# Informational and entropic criteria of self-organization

Zeinulla Zh. Zhanabaev[a], Yeldos T. Kozhagulov[a,*], Serik A. Khokhlov[a]

[a]Al-Farabi Kazakh National University, al-Farabi Avenue, 71, Almaty, Kazakhstan 050040 Tel. 8 (727) 221-15-52, Fax +7 (727) 377-33-44. kazgu.kz@gmail.com.

ABSTRACT

The work is devoted to study of the following problem: can we use any qualitative criteria for realization of such universal phenomenon as self-organization in open systems?

We have defined values of information at fixed points of probability function of density of information $I_1$ and entropy $I_2$. Physical meaning of these values as criteria of self-affinity and self-similarity in chaotic processes have been explained.

We have shown that self-organization occurs if normalized information entropy S belongs to the range $I_1 \leq S \leq I_2$, where $I_1 = 0.567$, $I_2 = 0.806$. The validity of these findings is confirmed by calculation of value of S for hierarchical sets of well-known fractals.

## 1. Introduction

Because of rapid development of modern technologies we need to know physical regularities of nanostructures, microwave chaotic signals, ensembles of neural networks, and so on. In spite of complexity of these objects and processes they have one common property which is the scale invariance. It means that for the description of such objects it is not necessary to use parameters with a concrete dimension, for example, dimension of length. Self-similarity (object characterized by only one value of similarity factor on different variables) and self-affinity (several similarity factors need for its description) are types of scale invariance. The common property of processes with different nature is self-organization of matter and its motion. Theory of self-organization is called the synergetics. Self-organization can be also considered as appearance of order from chaos if described system is open, nonlinear and non-equilibrium. So, for simplicity we shall discuss invariant properties of chaotic processes.

As usual, informational entropy and fractal dimension of a set of physical values are used as quantitative characteristics of chaos [1, 2]. According to the well-known I. Prigozhin theorem, the derivative of entropy with respect to time tends to its minimal value at self-organization in a system. According to the Yu. Klimontovich theorem [3], entropy of a system decreases at self-organization if energy of system is constant. Entropy can also be applied to guided self-organization [4].

However, accurate calculation of the direct normalized entropy of heterogeneous object is not realized, mentioned theorems do not answer the questions: what is the minimum value of entropy production, how entropy decreases at self-organization? It is also not clear the relation between entropy criterion of self-similarity and self-affinity and the fractal dimension characterizing corresponding chaotic processes. The purpose of this work is to search for answers to these questions.

## 2. Information criteria of scale invariance

Conception of information widely used in cybernetics, genetics, sociology, and so on. Development of synergetics and physics of open systems stimulates formulation of a universal definition of information which can be used in different branches of science. The definition of open system contains the conception of information: open system is a system in which energy, matter and information are exchanged with its environment.

As usual, definition of a complex object includes its main properties. Information $I(x)$ for statistical realization of a physical value $x$ is greater than zero and can be defined in non-equilibrium state ($I(x) \neq I(x_0)$, if $x \neq x_0$). Let us consider that $P(x)$ is probability of realization of a variable $x$. So, expression for the description of quantity of information can be written as

$$I(x) = -\ln P(x). \qquad (1)$$

Reiteration and non-equilibrium character of a process can be taken into account by the condition $0 < P(x) < 1$. A lot of definitions of information have been suggested in different branches of science, but Eq. (1) corresponds with all of them.

Information can be defined as

$$I(x/y) = S(x) - S(x/y), \qquad (2)$$

where $S(x)$ is absolute information entropy of an event $x$ and $S(x/y)$ is conditional entropy of an event $x$ when another event $y$ is to have occurred. The Eq. (2) can be used for solving of technical problems such as for estimation of transmission capacity (in communication channels). Informational entropy which is Shannon entropy $S(x)$ can be defined as mean value of information as

$$S(x) = \sum_i P_i(x) I_i(x) = -\sum_i P_i(x) \ln P_i(x). \qquad (3)$$

Here, $i$ is number of a cell after segmentation of $x$. So, let us use the Eq. (1) as a main definition of information.

According to Eq.(3) entropy calculated via probability density tends to infinity if $x$ is a continuous value. We try to define scale-invariant regularities. So, we must use a new approach for the description of information phenomena. Because of this fact we can use information as a defining independent variable. Statistical characteristics of a process can be described via information. So, we can try to find new properties of information independent on scale of measurement.

Therefore, according to Eq. (1) we shall describe probability of realization of information $P(I)$ as

$$P(I) = e^{-I}. \qquad (4)$$

Probability function $f(I)$ can be defined via the following relations:

$$\begin{aligned} & 0 \leq P(I) \leq 1, 0 \leq I \leq \infty, \int_0^\infty f(I) dI = 1, \\ & P(I) = \int_I^\infty f(I) dI, \ f(I) = P(I) = e^{-I}. \end{aligned} \qquad (5)$$

Probability function $P(I)$ equals to density of probability distribution function $f(I)$. Information defined via Eq. (1) characterized by property of scale invariance. It means that a whole object and its part have the same law of distribution. Informational entropy $S(I)$ of distribution of information can be defined as mean value of information as

$$S(I) = \int_I^\infty I f(I) dI = (1 + I) e^{-I}. \qquad (6)$$

Values of entropy can be normalized to unit, so, for $0 \leq I \leq \infty$ we have $1 \geq S \geq 0$. It is well-known that entropy of a continuous set tends to infinity at jumping values of variables. Therefore, we must take the integral by use of Lebesgue measure. By choosing information as the measure we have got Eq. (6).

Let us consider that a scale-invariant function $g(x)$ justifies to the well-known functional equation as

$$g(x) = \alpha g(g(x/\alpha)), \qquad (7)$$

where $\alpha$ is a scaling factor. All continuous functions in theirs fixed points justify the Eq. (7). We use $f(I)$ and $S(I)$ as characteristic functions. Fixed points of the functions are [5]:

$$f(I) = I, \ e^{-I} = I, \ I = I_1 = 0.567, \qquad (8)$$

$$S(I) = I, \ (1 + I)e^{-I} = I, \ I = I_2 = 0.806. \qquad (9)$$

The fixed points are limits for the following infinite maps

$$I_{i+1} = f(I_i), \ \lim_{i \to \infty} exp(-exp(\ldots - exp(I_0) \ldots)) = I_1, \qquad (10)$$

$$I_{i+1} = S(I_i), \lim_{i \to \infty} exp(-exp(... - exp(\ln(I_0 + 1) - I_0)...)) = I_2, \quad (11)$$

at any initial values $I_0$. Number of brackets equals to $i + 1$.

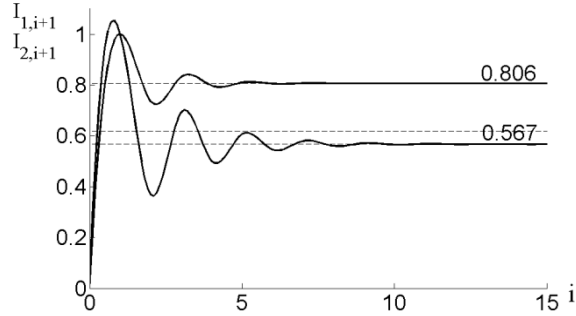

**Fig. 1.** Establishment of information self-similarity and entropy.

Interpretations of physical meaning of numbers $I_1 = 0.567$ and $I_2 = 0.806$ can be different. Probability density is a local (instant) characteristic. Therefore, it can be different for different variables. So, the number $I_1$ can be used as a criterion of self-affinity. Entropy is an averaged characteristic. So, value of $I_2$ is the criterion of self-similarity.

On the other hand, numbers $I_1$ and $I_2$ can be considered as analog of the Fibonacci number $I_{20} = 0.618$ ("golden section" of dynamical measure) for statistical self-affine and self-similar systems correspondently. From Eq. (9) at $I \lesssim 1$ we have

$$(1 + I)(1 - I) = I, \ I^2 + I - 1 = 0, \ I = I_{20} = 0.618, \quad (12)$$

at $I \ll 1$ from the same equation we have $e^{-I} = I$, $I = I_1$. Therefore, we can use only Eq. (9) for the description of regularities of self-affinity, dynamical equilibrium and self-similarity of dynamical systems.

## 3. Normalized information entropy of heterogeneous sets

Criteria of self-organization which we have established for nonequilibrium processes (heterogeneous objects) can be differ from $I_1, I_2$. It's necessary to consider influence of some parameter taking into account the deviation from equilibrium of the objects as a whole.

In recent years the new generalized statistical mechanics has been developed, which can be called the statistics of Tsallis [6], or quasicanonical Gibbs distribution [7, 8]. At the basis of such theories is using of exponential function, for example:

$$exp_{q-1}[x] = (1 + (q - 1)x)^{\frac{1}{q-1}}, \quad (13)$$

where $q$ is the degree of homogeneity, incompleteness parameter of statistical ensemble. In the limit $q \to 1$, we get the usual exponential. Within the meaning of entering

$$|q - 1| \sim \frac{1}{M}, M = L - N, M \to \infty, q \to 1, \quad (14)$$

where $L$ is the number of particles of the closed system, $N$ is the number of particles of the subsystem. The completeness of statistics corresponding to the canonical Gibbs's distribution of equilibrium is reached at $q = 1$. The distinction of unit parameter $q$ characterizes the degree of statistical non-equilibrium and the heterogeneity of the system. Nonequilibrium state of a physical value under examination can be taken into dividing of this value to intervals.

We define the full entropy $S(x, y)$ according to given degree of homogeneity of an object. Let us use as variables the one-dimensional and conditional probabilities $z_1 = P(x), z_2 = P(y/x)$. According to Eq. (13) we have

$$exp_{q-1}[z_1]exp_{q-1}[z_2] = exp_{q-1}[z_1 + z_2 + (q-1)z_1z_2]. \tag{15}$$

Expressing the left side as "logarithm $q-1$" of the product, we obtain

$$\ln_{q-1} z_1 z_2 = \ln_{q-1} z_1 + \ln_{q-1} z_2 + (q-1)\ln_{q-1} z_1 \ln_{q-1} z_2. \tag{16}$$

From Eq. (16) followed the expression for the non-additive "$S_{q-1}$ −entropy":

$$S_{q-1}(x,y) = S_{q-1}(x) + S_{q-1}(y/x) + (q-1)S_{q-1}(x)S_{q-1}(y/x). \tag{17}$$

In the limit of $q \to 1$ we have additive entropy $S(x,y) = S(x) + S(x/y)$.

According to the definition and Eq. (14) value of $q$ can be determined from experimental data. In order to describe the heterogeneity of geometric objects we enter the parameter $q \approx 1$. The algorithm for determining $q$ can be accepted as

$$q = \frac{N + <m>n(\delta)}{N} = 1 + \frac{<m>n(\delta)}{N} = 1 + \varepsilon, \quad 0 < \varepsilon < 1, \tag{18}$$

where $N$ is total number of points (samples), $n(\delta)$ is number of cells with the scale of measurement equal to $\delta$ containing at least one point, $<m>$ is average number of points in the cell.

Using the expression (17), we can define dependence of information entropy on $q$. We consider entropy as a single measure of complexity and uncertainty of non-equilibrium system. For a quasi-equilibrium process characterized by the parameter $q$, information can be defined as

$$I = -\ln_{q-1} P. \tag{19}$$

Thereafter, via the formulas (5), (6) and (13) we can use the expressions for probability density implementation and information entropy:

$$f_q(I) = exp_{q-1}[-I] = (1 - (q-1)I)^{\frac{1}{q-1}}. \tag{20}$$

$$S_q(I) = \int_I^\infty I f(I) dI = (1+I)exp_{q-1}(-I) = (1+I)(1-(q-1)I)^{\frac{1}{q-1}}. \tag{21}$$

These expressions contain parameter q. Self-similar values $f(I_1) = I_1$ and $S(I_2) = I_2$ we can define as fixed points of maps

$$I_{1q,i+1} = (1-(q-1)I_{1,i})^{\frac{1}{q-1}}, \tag{22}$$

$$I_{2q,i+1} = (1+I_{2,i})(1-(q-1)I_{2,i})^{\frac{1}{q-1}}, \tag{23}$$

$$I_{1q,0} = I_{2q,0} = 0, \infty; \quad i = 0,1,2....$$

Thus, value of $q$ can characterize deviation of the state of the system from self-similarity and self-affinity via values of information and informational entropy.

Meaning of the parameter q is expanding even more, if we take into account the well-known expression for the multi-fractal dimension $D_q$ and Renyi entropy $S_R$ containing a parameter with the same designation

$$D_q = \lim_{\delta \to 0} \frac{1}{q-1} \frac{\ln \sum_i p_i^q}{\ln \delta}, \quad S_R = -\frac{\ln \sum_i p_i^q}{q-1} \tag{24}$$

At $q \to 1$ entropies of Renyi, Tsallis (17) transform to the Shannon's entropy (3). Therefore, at $q = 1 + \varepsilon(\delta) \succ 1$ this parameter can be considered as fractional order of multifractal moment. It is important that this parameter can be determined from observations (formula (18)).

Universal entropic regularities of evolution of open systems to self-similarity and self-affinity modes according to Eq. (22) and (23) are shown in Fig. 2, where we accept characteristic time as $t = i$. A fluctuation from equilibrium leads to qualitatively different patterns of establishing the scale invariance for $q < 1$ and $q > 1$ (Fig. 2).

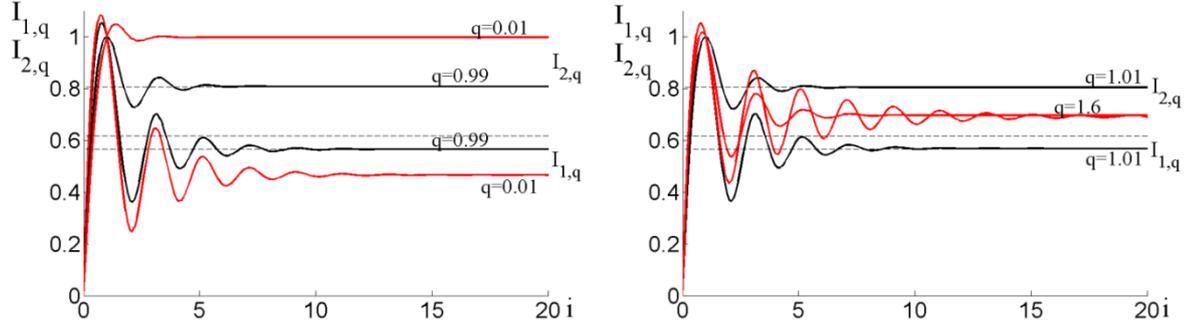

**Fig. 2.** Establishing the evolution of information ($I_{1,q}$) and entropy ($I_{2,q}$) for different '$q$'.

## 4. The results of numerical analysis and its consideration.

At first, we note the possibility for describing various types of entropy and dimension of fractal sets by use of parameter $q$ in the form (18). For this aim we shall define $q$ for the sets of hierarchy of fractals which are pre-fractals of $n$ generation. Then we have a possibility to define value of multifractal dimension $D_q$. In this case the algorithm for definition $D_q$ will be different from the standard algorithm: it is necessary to take into account dependence of $q$ on size of a cell $\delta$ via $\varepsilon = \varepsilon(\delta)$. Because of this the dependence $\ln \sum_i p_i^q$ on $\varepsilon(\delta)\ln(\delta)$ clearly separated for each generation of $n$-pre-fractal, and corresponding values $D_{q,n}$ are different (Fig. 3). So, we rise the problem for selecting of value $D_{q,n}$ for normalization of entropy. To establish the criterion of self-organization it is necessary to use value of $D_q$ determined by the greatest interval of scale invariance of $\varepsilon(\delta)\ln(\delta)$ variable. $[\delta_1, \delta_2]$ interval for rectilinear portion in Fig. 3 for $n$-pre-fractal

$$\Delta(n) = \varepsilon_n(\delta_2)\ln\delta_2 - \varepsilon_n(\delta_1)\ln\delta_1 \tag{25}$$

should be maximal. Indeed, as follows from Fig. 3, where, and further, the names of fractals correspond to the references [9, 10] has maximum at $n \succ 3$ (Fig. 4). It means that after reaching the maximum interval of self-similarity the further increasing of pre-fractal order leads to refinement of its structure. Therefore, minimum value $\Delta(n)$ corresponds the transition from self-similarity to self-affinity and multifractality with $q_n = 1 + \varepsilon_n$. Values of Hausdorff dimension $D_{0,n} (D_0 = D_{q=0})$ calculated for $n$-pre-fractals and multifractal dimension $D_{q,n}$ approximate at selecting the minimum value $\Delta(n)$ (Fig. 5) at $n \geq 3$.

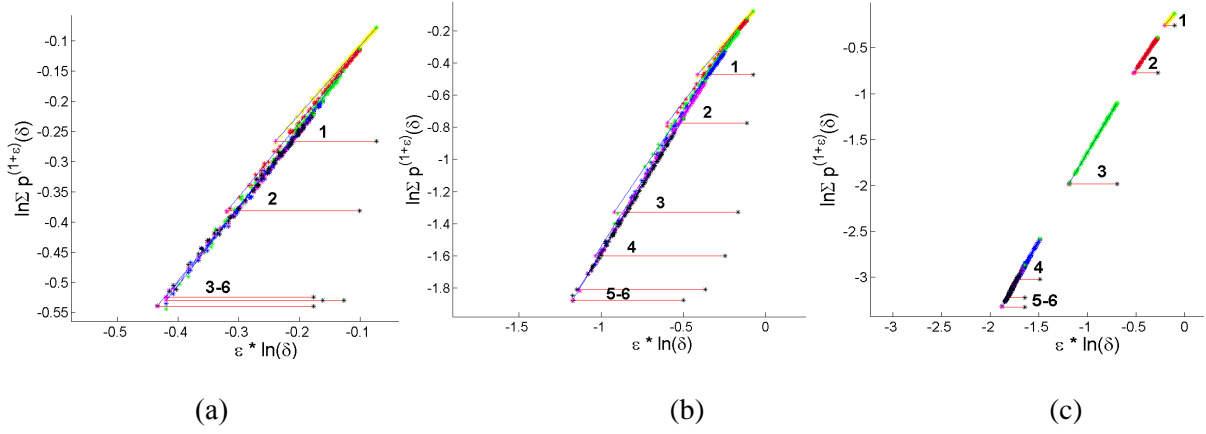

**Fig. 3.** Different intervals of self-similarity for $n$-pre-fractals: a) Koch curve, b) Sierpinski curve, c) Mandelbrot-Given curve. $1 \leq n \leq 6$ values shown on the figure.

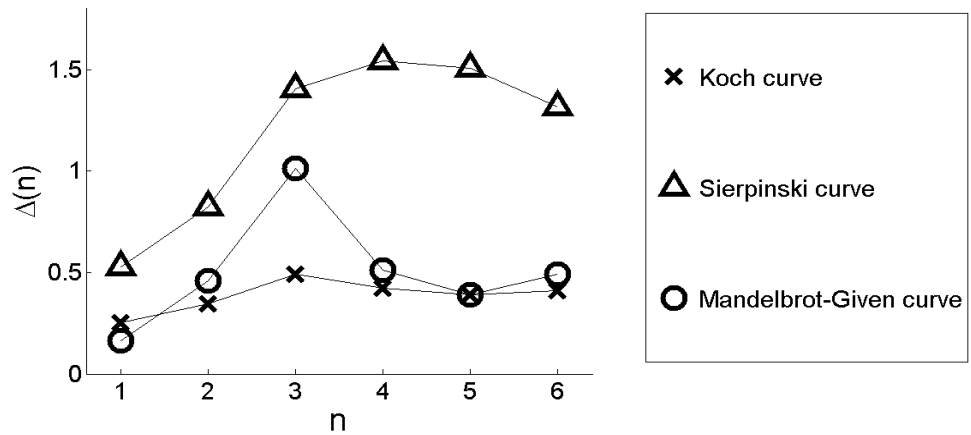

**Fig. 4.** Dependence of length on self-similarity interval of pre-fractal number.

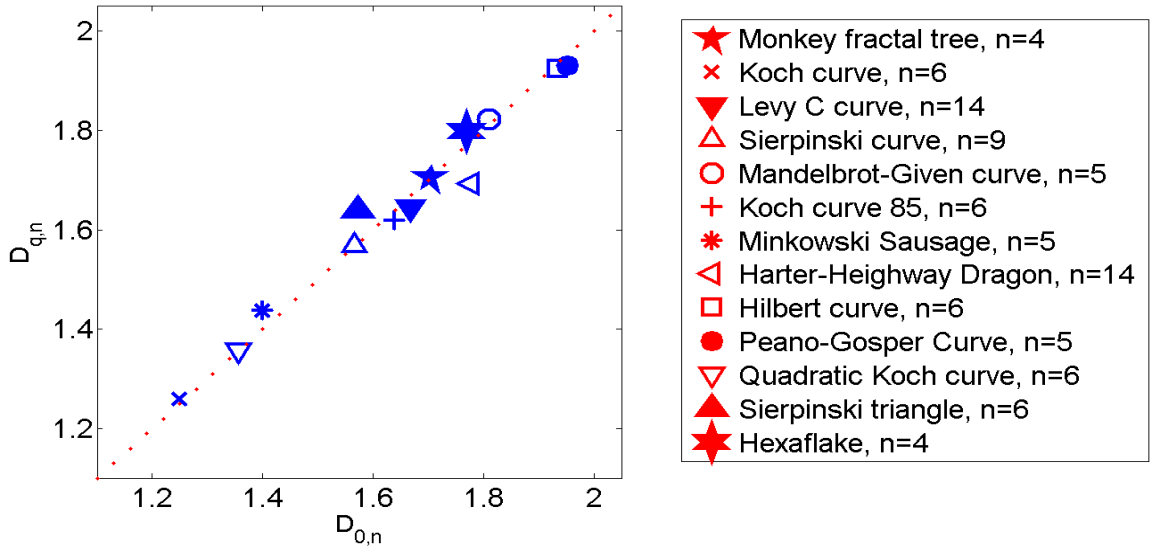

**Fig. 5.** Equivalence of multifractal dimension $D_{q,n}$ of $n$-pre-fractals for $\Delta_{\min}(n)$ with the corresponding Hausdorff dimensions $D_{0,n}$.

An interval of values of normalized Shannon entropy $I_1 \leq S/S_{max} \leq I_2$ defines a self-organized process, including its modes of self-similarity and self-affinity. It follows from Figure 2 for the case $q > 1$. This fact facilitates the physical analysis of random processes, if the way to the normalization of entropy is known. For one-dimensional signal $S_{max}$ is equal to entropy of isosceles triangular pulse, distribution of points by the step of argument is equable in this case.

For two-dimensional realizations method for normalization of entropy is the problem because there is no known method of selecting samples of identically distributed readings of two variables of different nature (different evolution). Entropy is a measure of the uncertainty of a system and increases with the degree of heterogeneity ($q-1=\varepsilon$ parameter). Consequently, we can use the Renyi entropy $S_R$ as the normalizing value defined by $q$ which is a moment of probability of realization of a physical value by the formula (24). If the interval of change of characteristic scale $\delta$ includes minimum and maximum values defining the variables of the system, $S_R$ should be regarded as its own entropy norm of the system. This approach can have universal applying, because open systems do not have the absolute norm.

Thereafter normalized Shannon entropy (formula (3)) is defined as $S = S(x)/S_R$. Normalization on Tsallis entropy (formula (17)) will be less exact, since "$q$-exponential" greatly varies at deviation $q$ from the unit. But the use of this approach has proved useful for analytical explaining the role of parameter $q$ to the value of self-similarity of the normalized entropy $S(I)$ (Fig. 2).

The conclusions of this work according to the criteria of self-organization can be checked by calculating the normalized information entropy of fractals according to the degree of their heterogeneity. Geometric fractals are convenient models of hierarchical dynamic system, because each generation of fractal can be associated with an evolutionary level.

Figure 6 shown the dependence of $S = S_n(x, y; \delta)/S_{R,n}$ on $D_{0,n}$ where $S_n(x, y; \delta)$ is Shannon entropy of $n$-pre-fractal on the plane $(x, y)$, $S_{R,n} = D_{q,n} \ln N(\delta)$ is the corresponding Renyi entropy defined by multifractal dimension $D_{q,n}$, and the number of filled cells of the partition $N(\delta)$ with characteristic scale $\delta$ within the interval of self-similarity, $D_{0,n}$ is Hausdorff dimension of $n$-pre-fractal. Hierarchical number of fractal $n$ corresponds to the maximum interval of self-similarity in the variable $\varepsilon \ln N(\delta)$. The normalized values of information entropy of all considered fractals are in the interval of self-organization $[I_1, I_2]$.

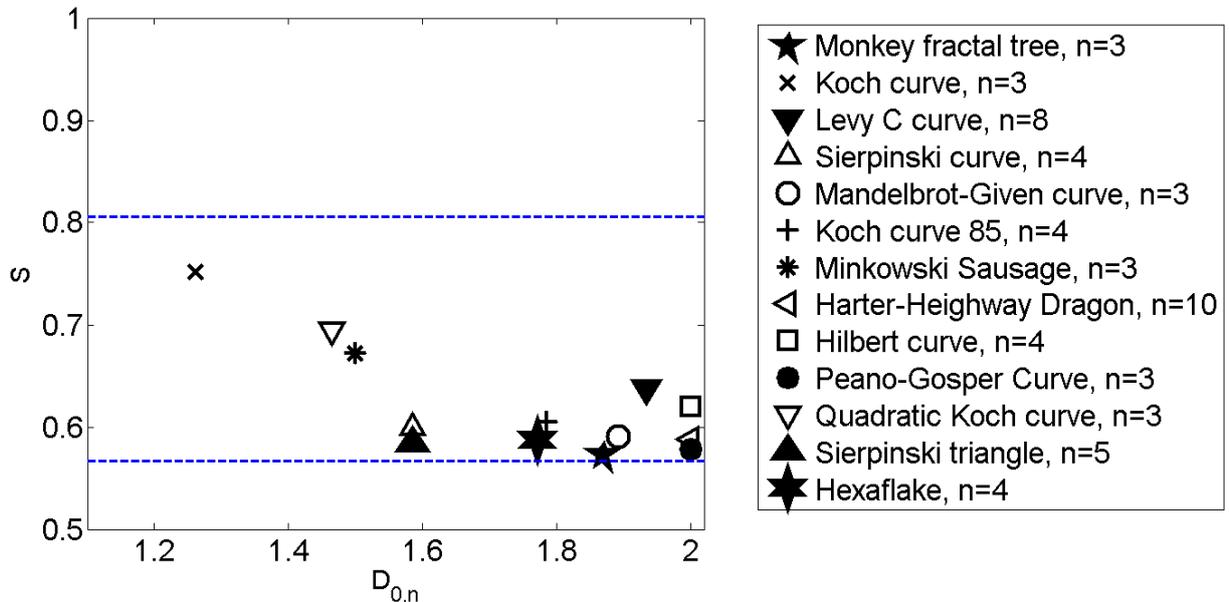

**Fig. 6.** Normalized information entropy of $n$-pre-fractals.

**5. Conclusion**

In this paper, we obtain an analytical expression for the dependence of normalized entropy from information. It is shown that the fixed points of entropy and density probability of realization of information determine an interval of self-organization, which includes self-similarity and self-affinity modes.

In order to use results of the theory to concrete open systems we accept norm of Shannon entropy as Renyi entropy related with multi-fractal dimension. We propose the algorithm for determining from observations order of multifractal moment. Results of the theory are confirmed by calculating the normalized entropy sets of hierarchies known fractals. These quantitative information-entropic criteria of self-organization and related algorithms of processing results of observations can be used for the description of open systems with different nature.